\documentclass[10pt]{article}
\usepackage[]{amsmath,amssymb}
\usepackage{amsfonts}
\usepackage{graphics,epsfig}
\def\ie{{\rm i.e.,\/}\ }
\def\etc{{\rm etc.\/}\ }

\def\one{\mbox{\rm 1}\hskip-2.8pt \mbox{\rm l}}

\newcommand{\ZZ}{\mathbb{Z}}
\newcommand{\RR}{\mathbb{R}}
\newcommand{\CC}{\mathbb{C}}

\title{Classical and quantum polyhedra:\\
A fusion graph algebra point of view 
}

\vspace{0.7 cm}

\author{R. Coquereaux${}^1$\thanks{~Email: coque@cpt.univ-mrs.fr},
\\
{\small Lectures given at the Karpacz winter school (February 2001)}\\
${}^1$ {\it Centre de Physique Th\'eorique - CNRS - Luminy, Case 907} \\
       {\it F-13288 Marseille Cedex 9 - France}}

\begin{document}
\begin{titlepage}

\maketitle


\abstract{
Representation theory, for the classical binary polyhedral groups 
$\tilde T$, $\tilde C$ and $\tilde I$,  is 
encoded by the affine Dynkin diagrams $E_6^{(1)}$, $E_7^{(1)}$ and 
$E_8^{(1)}$ (McKay correspondence). 
The quantum versions of these classical geometries are associated with representation 
theories  described by the usual Dynkin diagrams $E_{6}$, $E_{7}$ 
and $E_{8}$.
The purpose of these notes is to compare several chosen aspects of the classical
and quantum geometries by using the study of spaces of paths and 
spaces of essential paths (Ocneanu theory) on these diagrams.
To keep the size of this contribution small enough, most of our discussion 
will be limited to the cases of 
diagrams $E_{6}$ and $E_{6}^{(1)}$, \ie to the classical and quantum 
tetrahedra. 
We shall in particular interpret the $A_{11}$ labelling of $E_{6}$ vertices 
as a quantum analogue of the usual decomposition of spaces of 
sections for vector bundles above homogeneous spaces. We  also 
show how to recover Klein invariants of polyhedra by paths algebra techniques 
and discuss their quantum generalizations.
}
\vskip 3cm

CPT-2001/P4208

hep-th/0105239
\end{titlepage}


\vfill\eject

\section{Introduction}

One way to understand the classification of modular invariant 
partition functions, for instance the $ADE$ classification of $SU(2)$ conformal 
theories (\cite{CIZ}, \cite{Pasquier}) or its generalizations 
(\cite{DiFrancescoZuber}, \cite{DiF-Zub-Trieste}), should
be understood, following A. 
Ocneanu (many talks since 1995, for instance \cite{Ocneanu:Marseille}, and \cite{Ocneanu:paths})  through the study of ``quantum symmetries'' on graphs, and in particular, on 
Dynkin $ADE$ diagrams. In turn, this study leads, in each case, to new kinds of 
partition functions wich are not modular invariant but are 
nevertheless quite remarkable: the concept of ``torus structure'' of ADE graphs 
is due to A.Ocneanu (unpublished), the corresponding ``twisted partition  
functions'' have been recently discussed by (\cite{PetkovaZuber:Oc}),
the expressions of  the``toric matrices'' for the quantum 
tetrahedron ($E_{6}$ model) have been presented
 in (\cite{Coquereaux:QuantumTetra}) and
explicit calculations for the other models, using the techniques of 
this last paper should appear in \cite{CoqueSchieber:Oc}. The needed 
mathematical material is unfortunately not really standard and often not even
available in published form. It happens, however, that many of these 
algebraic (``quantum'') manipulations can be seen as a quantum analogue of finite 
group constructions. The purpose of these notes is to 
give a fresh look to the old-fashion representation theory of groups of 
symmetries of regular polyhedra, and to do it in such a way that 
generalization to the quantum case becomes (almost) straightforward.
However, we shall not cover all the way, from $ADE$ 
diagrams to Ocneanu quantum symmetries, even when discussing the 
classical analogue of these constructions: in particular the study of 
Racah-Wigner bi-algebras for Platonic groups will be left aside 
(this should be done within \cite{CGT:triangles}) and we shall 
not even introduce the classical analogue of Ocneanu graphs\ldots
In particular, the twisted partition functions that we can associate
to the different vertices of the Ocneanu graph will not be described.
We hope nevertheless that this set of notes will provide a useful
introduction to this fascinating subject.

The major part of what is going to be explained below is certainly known, at 
least in some circles. We believe, however, that our 
geometrical interpretation, for
the ``essential'' labelling of  vertices of exceptional Dynkin diagrams
 (like $E_{6}$) by  vertices belonging to an
appropriate $A_{N}$ graphs (which is  $A_{11}$ in the case of 
$E_{6}$), as the quantum analogue of the decomposition of a space of 
sections of a homogeneous vector bundle, 
is probably new, and may be of interest for the  expert. 
Also, explicit calculations of projectors decomposing the representations 
$[2]^p$ of binary polyhedral groups into irreps have been probably
carried out by several group theorists, chemists, or solid state physics practitioners, 
but we do not think that a description  of the  method
using paths on graphs together with the data provided by  spectral 
decomposition of the (quantum) $\hat R$ matrices 
was given before.
One should be aware, however, that in several cases, like for instance 
 the calculation of Klein 
invariants,  the results themselves have been known for more  than a century!

\section{Classical and quantum Platonic bodies}

As it is well known, geometry of classical Platonic bodies  is
encoded by their symmetry groups that we shall call respectively $T$ 
(for the the tetrahedron), $C$ (for the cube and its dual, the 
octahedron) and $I$ (for the icosahedron and its dual, the 
dodecahedron). These are subgroups of $SO(3)$. By adding reflections, we may also
consider the ``full'' polyhedral groups, which are subgroups of
$O(3)$, but the objects of interest, for us, are the so-called binary
polyhedral groups $\tilde T, \tilde C, \tilde I$,
that are non abelian subgroups of $SU(2)$, the two-fold cover of $SO(3)$.
They are defined as pre-image of the corresponding $SO(3)$ subgroups by using the sequence
$$ 0 \mapsto \ZZ_2 \mapsto SU(2) \mapsto SO(3) \mapsto 0$$

We concentrate our attention mostly on the particular example given by the binary {\sl tetrahedral} group (of order
$2\times 12 = 24$) since we want to compare it, or better its representation theory, with a kind
of quantum analogue. One reason to limit our study to $\tilde T$ is lack 
of space, the other is that the quantum geometry corresponding to $C$ 
is somehow much more difficult to study that the one corresponding 
to $T$ and $I$.

\subsection{Classical geometry}

\subsubsection{Several realisations of the group $\tilde T$}

Elementary geometric considerations show that $T$ itself is 
isomorphic with the group of even permutations on four elements 
(label the vertices of the tetrahedron by $(1,2,3,4)$). Therefore
$$\# \tilde T = 2 \times \# T = 2 \times \frac{4!}{2} = 2 \times 12 = 
24$$

Since $\tilde T$ is, by definition, a finite subgroup of 
$Spin(3) (\simeq SU(2))$, we can express all its elements in terms of 
$Cliff(\RR^{3})$, \ie in terms of the ``gamma matrices'' of $\RR^{3}$, 
namely the Pauli matrices $\tau_{i}$. Setting $\gamma_{x}\doteq - 
\tau_{2}$, 
$\gamma_{y}\doteq\tau_{1}$ and $\gamma_{z}\doteq \tau_{3}$, we get
$\gamma_{i}\gamma_{j}+\gamma_{j}\gamma_{i} = 2 \delta_{ij}$ with
$$
\begin{array}{ccc}
\gamma_{x} = 
\left( 
\begin{array}{cc}
0 & i \\
-i & 0  
\end{array}
\right)
&
\gamma_{y} = 
\left( 
\begin{array}{cc}
0 & 1 \\
1 & 0  
\end{array}
\right)
&
\gamma_{z} = 
\left( 
\begin{array}{cc}
1 & 0 \\
0 & -1  
\end{array}
\right)
\end{array}
$$
It is easy to see that $\tilde T$ is generated, as a group, by
\begin{eqnarray*}
{\bf t} &=& \gamma_{x} \gamma_{y} \cr
{\bf s} &=& \frac{1}{2} (\gamma_{x} \gamma_{y} + \gamma_{y} \gamma_{z} + 
\gamma_{z} \gamma_{x} - 1)
\end{eqnarray*}
Notice that ${\bf t}^{2}=-1, {\bf t}^{3}=-{\bf t},{\bf t}^{4}=1$ and ${\bf s}^{2}=-({\bf s}+1)$, so that 
${\bf s}^{3}=1$. Moreover $({\bf s}{\bf t})^3=1$.
Explicitly:
$$
\begin{array}{cc}
{\bf s} \doteq \left( 
\begin{array}{cc}
(-1+i)/2 & (-1+i)/2 \\
(1+i)/2 & -(1+i)/2  
\end{array}
\right)
&
{\bf t}  \doteq \left( 
\begin{array}{cc}
i & 0 \\
0 & -i 
\end{array}
\right)
\end{array}
$$

The reader may  be interested in knowing that $\tilde T$ (resp. $\tilde C$ and $\tilde I$) is isomorphic
with the group $\langle 2,3,n \rangle$ (Threlfall 
notation), when $n=3$ (resp. $n=4$ and $n=5$). 
This notation refers, by definition, to the group generated by two elements $A$ and $B$, with
relations $A^3=B^n=(AB)^2$. These groups, when $n=3$ or $n=5$ are also isomorphic
with the groups $SL(2,F_{3})$ or $SL(2,F_{5})$, here $F_{n}$ is the field with $n$ 
elements; there is no such isomorphism  when $n=4$.
 It may be also nice to 
remember that the ``extended'' triplets  $\langle 3,3,3 \rangle$,  $\langle 2,4,4 
\rangle$ and $\langle 2,3,6 \rangle$ are the only positive integer solutions of the equation :
$$
\frac{1}{a} + \frac{1}{b} + \frac{1}{c} = 1
$$
These extended triplets encode the affine Dynkin diagrams of type $E_{6}^{(1)}$, 
$E_{7}^{(1)}$, $E_{8}^{(1)}$ (lengths of the three legs counted from the 
triple point).
Substracting $1$ from a maximal element of these extended triplets give the  
triplets  $\langle 2,3,3 \rangle$,  $\langle 2,3,4
\rangle$ and $\langle 2,3,5 \rangle$ which satisfy the 
inequality\footnote{The other solutions of this inequality, if we exclude $a=1, 
b\neq c$, are $\langle 1,n,n \rangle$ and $\langle 2,2,n \rangle$, 
corresponding to the $A_{2n-1}$ and $D_{n+2}$ Dynkin diagrams.}
$$
\frac{1}{a} + \frac{1}{b} + \frac{1}{c} > 1
$$
and encode the usual Dynkin diagrams for exceptional Lie groups. The 
same triplets also characterize the binary polyhedral groups since 
they give the relations obeyed by their generators in Threlfall 
notation.

\subsubsection{Representations of the group $T$}
One way to get the representations of $T$ is to remember that $T \simeq A_4 \subset S_4$
and use the fact that representation
theory of permutation groups (like $S_4$) is well known (use Young tableaux, for instance).

$T\simeq A_4$ has four irreducible inequivalent representations of respective dimensions $1,1,1,3$.
We call them $\sigma_1$,
$\sigma_1'$, $\sigma_1''$, $\sigma_3$. We check that the dimensions divide $12$ (as they should) and that $1^2+1^2+1^2 + 3^2
= 12$, of course.
\subsubsection{Representations of the group $\tilde T$}
The previous irreps are also irreducible representations for the corresponding binary group
$\tilde T \subset SU(2)$, but
the latter also possesses other irreducible representations.
Altogether, $\tilde T$ has
seven irreducible inequivalent representations; their dimensions are $$1,1,1,3, 2, 2, 2.$$ The odd dimensional ones will be
labelled as before  and the last three will be called $\sigma_2$, $\sigma_2'$ and $\sigma_2''$. The one called $\sigma_2$ is,
by definition, the fundamental of $SU(2)$ restricted to $\tilde T$.
With no surprise we check that all these dimensions divide $24$ and that $1^2+1^2+1^2 + 3^2 + 2^2 + 2^2 + 2^2 = 24$.
The representations $\sigma_1$, $\sigma_2$, $\sigma_3$ are 
self-conjugate (actually real) $\sigma_1'$,$\sigma_1''$, 
$\sigma_2'$,$\sigma_2''$ are complex and respectively conjugated to 
one another.
\subsubsection{Tensor products of representations and McKay  correspondence}
In the case of $SU(2)$, it is well known that the tensor product of the representation of spin $1/2$ (\ie dimension $2$) by
a representation of spin $j$ (dimension $n=2j+1$) is equivalent to the sum of two representations of respective spin $j -1/2$
and $j+1/2$. In terms of dimensions, we have $2 (2j+1) = (2j)+(2j+2)$; in terms of the representations themselves (working up
to equivalence) we have: $$\sigma_2 \otimes \sigma_{n} = \sigma_{n-1} \oplus \sigma_{n+1}$$
The irreps into which a given representation $\sigma_n$ decomposes, upon tensorial multiplication by the
fundamental representation $\sigma_2$ are given by the neighbours of 
$\sigma_n$ on the  following semi-infinite diagram
 called the $A_{\infty}$ diagram.

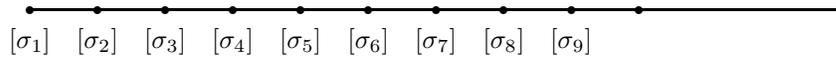
\begin{figure}[htb]
\unitlength 0.6mm
\begin{picture}(95,35)
\thinlines 
\multiput(25,10)(15,0){10}{\circle*{2}}
\thicklines
\put(25,10){\line(1,0){180}}
\put(25,3){\makebox(0,0){$[\sigma_1]$}}
\put(40,3){\makebox(0,0){$[\sigma_2]$}}
\put(55,3){\makebox(0,0){$[\sigma_3]$}}
\put(70,3){\makebox(0,0){$[\sigma_4]$}}
\put(85,3){\makebox(0,0){$[\sigma_5]$}}
\put(100,3){\makebox(0,0){$[\sigma_6]$}}
\put(115,3){\makebox(0,0){$[\sigma_7]$}}
\put(130,3){\makebox(0,0){$[\sigma_8]$}}
\put(145,3){\makebox(0,0){$[\sigma_9]$}}
\end{picture}
\caption{The diagram $A_\infty$}
\label{graphAinfty}
\end{figure}

Returning to the binary tetrahedral group $\tilde T$, we decide to encode in the same way the tensor product of the
various irreps by the fundamental (the $2$-dimensional).
The calculation itself is a simple exercise in finite group theory
and we shall not dwell on the matter... The point is that, if we decide to encode the results in terms of a graph with seven
vertices (the seven irreps), this graph is nothing else than the Dynkin diagram of the exceptional affine Lie algebra
$E_6^{(1)}$.

What comes as a surprise is that, if we perform the same construction with the binary groups of the cube and of the
icosahedron, we obtain respectively the Dynkin diagrams of the affine Lie algebras  $E_7^{(1)}$ and $E_8^{(1)}$.
This observation is known as ``McKay correspondence'' (\cite{McKay}).

The diagrams $E_6^{(1)}$ and $E_8^{(1)}$,
labelled by the seven (nine) irreducible representations of the binary 
tetrahedral (icosahedral)
group are displayed below.

\bigskip

\begin{figure}[htb]
\unitlength 0.6mm
\begin{picture}(95,35)
\thinlines 
\multiput(10,10)(15,0){5}{\circle*{2}}
\put(40,25){\circle*{2}}
\put(40,40){\circle*{2}}
\thicklines
\put(10,10){\line(1,0){60}}
\put(40,10){\line(0,1){30}}
\put(10,3){\makebox(0,0){$[\sigma_1]$}}
\put(25,3){\makebox(0,0){$[\sigma_2]$}}
\put(40,3){\makebox(0,0){$[\sigma_3]$}}
\put(55,3){\makebox(0,0){$[\sigma_{2}']$}}
\put(70,3){\makebox(0,0){$[\sigma_{1}']$}}
\put(48,27){\makebox(0,0){$[\sigma_{2}'']$}}
\put(48,42){\makebox(0,0){$[\sigma_{1}'']$}}
\thinlines 
\multiput(85,10)(15,0){8}{\circle*{2}}
\put(160,25){\circle*{2}}
\thicklines
\put(85,10){\line(1,0){105}}
\put(160,10){\line(0,1){15}}
\put(85,3){\makebox(0,0){$[\sigma_1]$}}
\put(100,3){\makebox(0,0){$[\sigma_2]$}}
\put(115,3){\makebox(0,0){$[\sigma_3]$}}
\put(130,3){\makebox(0,0){$[\sigma_4]$}}
\put(145,3){\makebox(0,0){$[\sigma_5]$}}
\put(160,3){\makebox(0,0){$[\sigma_6]$}}
\put(175,3){\makebox(0,0){$[\sigma_{4}']$}}
\put(190,3){\makebox(0,0){$[\sigma_{2}']$}}
\put(168,27){\makebox(0,0){$[\sigma_{3}']$}}
\end{picture}
\caption{The diagrams ${E_6}^{(1)}$ and ${E_8}^{(1)}$}
\label{graphsE6E8affine}
\end{figure}
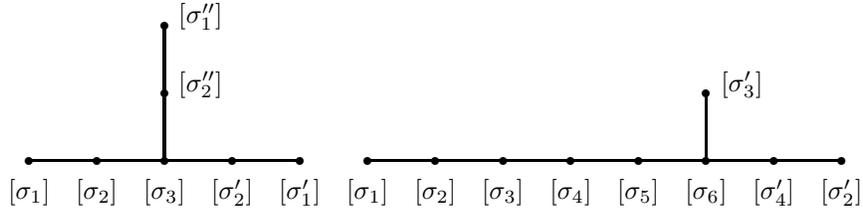

As explained before,
this reads, for instance  in the case of $E_6^{(1)}$,
 $\sigma_1' \otimes \sigma_2 = \sigma_2'$, or
$\sigma_2 \otimes \sigma_3 = \sigma_2 \oplus \sigma_2' \oplus \sigma_2'' $. 
The summands appearing on the right hand side are the neighbours, on 
the graph, of the chosen vertex.
Of course, the dimensions should match (so, for
instance $2\times 3 = 2 + 2 +2$).

For the above reason we decide to denote by ${\cal H}_{E_6^{(1)}}$, ${\cal H}_{E_7^{(1)}}$ 
and ${\cal H}_{E_8^{(1)}}$ the group algebras of these three binary groups. These 
are co-commutative finite dimensional semi-simple Hopf algebras. The 
corresponding dual objects (algebras of complex valued functions of these 
groups) are denoted by ${\cal F}_{E_6^{(1)}}$, ${\cal F}_{E_7^{(1)}}$ 
anf ${\cal F}_{E_8^{(1)}}$.

\subsubsection{Structure of the Grothendieck ring}

We already know how to multiply representations of $\tilde T$ by $\sigma_1$ (the trivial representation) and by $\sigma_2$
(this is given by the graph $E_6^{(1)}$). In other words, we know the first two rows and the first two columns of the
table of multiplication of characters.  Another useful identity
that one should get first is $1' 1'' = 
1$; this is a consequence of the fact that $1' 1''$ is one 
dimensional (so it is either $1$, $1'$ or $1''$) but it is real, since 
${\overline {1'}} = {1''}$.
This data is sufficient to reconstruct the whole table
of multiplication. We only have to use associativity of $\otimes$ and perform the calculations in the ring of virtual
representations (hence allowing minus signs at intermediate steps in our calculations.)

Let us for instance compute $\sigma_3 \otimes \sigma_3$ (we shall  just write $3 \,  3$ in the sequel):
\begin{eqnarray*}
3 \, 3 &=& 3 (2 2-1) = 3 2 2 - 3 = (3 2) 2 - 3 = (2 + 2' + 2'') 2 - 3 = \cr
&=& 1 + 3 + 1' + 3 + 1'' + 3 - 3 = 3 + 3 + 1 + 1' + 1'' = (3)_2 +  1 + 1' + 1''
\end{eqnarray*}

Here, the subindex $2$ in $(3)_2$ means that $3$ appears with multiplicity $2$.

After some work, all other entries of the multiplication table can be worked 
out by simple manipulations analoguous to the above  calculation of $3 
\, 3$ and  we obtain the following table (that we shall sometimes call the 
fusion table);  it gives all the coupling constants $C_{ijk}$ of 
the Grothendieck ring $Ch(G)=\ZZ(Irr(\tilde T))$; these constants are 
defined by
$\sigma_i \otimes \sigma_j = C_{ijk} \sigma_k$.

$$
\begin{array}{||c||c|c|c|c|c|c|c||}
\hline
{} & 1 & 2  & 1' & 2' & 1'' & 2'' & 3  \\
\hline
\hline
1 & 1 & 2  & 1' & 2' & 1'' & 2'' & 3 \\
2 & 2  & 1,3 & 2' & 1'3 & 2'' & 1''3 & 2 2' 2'' \\
1' & 1'  & 2' & 1'' & 2'' & 1 & 2 & 3 \\
2' & 2'  & 1'3 & 2'' & 1'' 3 & 2 & 1 3 & 2 2' 2''  \\
1'' & 1''  & 2'' & 1 & 2 & 1' & 2' & 3 \\
2''& 2''  & 1''3 & 2 & 1 3 & 2' & 1' 3 & 2 2' 2'' \\
3 & 3  & 2 2' 2'' & 3 & 2 2' 2'' & 3 & 2 2' 2'' & 1 1' 1'' 3_2
\\
\hline
\end{array}
$$

Notice that the final table only involves sums (no minus signs, of 
course!) of irreducible representations.

We should maybe write this table in the order $(1,1',1'',3 ; 2,2',2'')$ 
to reflect the fact that the subset $1,1',1'',3$ (irreps of $T$) is
stable under tensorial multiplication;  
this corresponds to the fact that the group $T$ is a quotient of $\tilde T$.

\subsubsection{The fusion matrices $N_{i}$}

The (non necessarily symmetric) matrices $N_{i}$ are 
defined by
$$
(N_{i})_{jk}= C_{ijk}
$$

Still using the order $\{1,1',1'',2,2',2'',3\}$,  we have :
{\small
$$
\begin{array}{ccc}

N_1 = \left( \begin{array}{ccccccc}
  1&0&0&0&0&0&0  \\ 0&1&0&0&0&0&0 \\
   0&0&1&0&0&0&0  \\ 0&0&0&1&0&0&0 \\
   0&0&0&0&1&0&0  \\ 0&0&0&0&0&1&0 \\
   0&0&0&0&0&0&1   \end{array} \right)

N_1' =
\left( \begin{array}{ccccccc}
0& 0& 1& 0& 0& 0& 0\\ 1& 0& 0& 0& 0& 0& 0\\ 0& 1& 0& 0& 0& 0& 
        0\\ 0& 0& 0& 0& 0& 1& 0\\ 0& 0& 0& 1& 0& 0& 0\\ 0& 0& 0& 0& 1& 0& 
        0\\ 0& 0& 0& 0& 0& 0& 1  \end{array}\right)

N_1'' = 
\left( \begin{array}{ccccccc}
0& 1& 0& 0& 0& 0& 0\\ 0& 0& 1& 0& 0& 0& 0\\ 1& 0& 0& 0& 0& 0& 
        0\\ 0& 0& 0& 0& 1& 0& 0\\ 0& 0& 0& 0& 0& 1& 0\\ 0& 0& 0& 1& 0& 0& 
        0\\ 0& 0& 0& 0& 0& 0& 1   \end{array}\right)

\end{array}
$$

$$
\begin{array}{ccc}

N_2 = 
\left( \begin{array}{ccccccc}
  0& 0& 0& 1& 0& 0& 0\\ 0& 0& 0& 0& 1& 0& 0\\ 0& 0& 0& 0& 0& 1& 0\\ 1& 0& 
    0& 0& 0& 0& 1\\ 0& 1& 0& 0& 0& 0& 1\\ 0& 0& 1& 0& 0& 0& 1\\ 0& 0& 0& 1&
     1& 1& 0  \end{array} \right)

N_2' = 
\left( \begin{array}{ccccccc}
0& 0& 0& 0& 0& 1& 0\\ 0& 0& 0& 1& 0& 0& 0\\ 0& 0& 0& 0& 1& 0& 
        0\\ 0& 0& 1& 0& 0& 0& 1\\ 1& 0& 0& 0& 0& 0& 1\\ 0& 1& 0& 0& 0& 0& 
        1\\ 0& 0& 0& 1& 1& 1& 0
   \end{array}\right)
   
N_2'' = 
\left( \begin{array}{ccccccc}
  0& 0& 0& 0& 1& 0& 0\\ 0& 0& 0& 0& 0& 1& 0\\ 0& 0& 0& 1& 0& 0& 
        0\\ 0& 1& 0& 0& 0& 0& 1\\ 0& 0& 1& 0& 0& 0& 1\\ 1& 0& 0& 0& 0& 0& 
        1\\ 0& 0& 0& 1& 1& 1& 0   \end{array}\right)

\end{array}
$$

$$
N_3= 
\left( \begin{array}{ccccccc}
0& 0& 0& 0& 0& 0& 1\\ 0& 0& 0& 0& 0& 0& 1\\ 0& 0& 0& 0& 0& 0& 1\\ 0& 0& 
    0& 1& 1& 1& 0\\ 0& 0& 0& 1& 1& 1& 0\\ 0& 0& 0& 1& 1& 1& 0\\ 1& 1& 1& 0&
     0& 0& 2
 \end{array}\right)
$$
}
It is easy to show that we obtain an isomorphism between 
the -- commutative --  ring of characters $Ch(\tilde T)$ (or, 
equivalently, the ring generated by tensor 
powers of the irreducible representations) and the ring generated (over 
the integers $\ZZ$) by the matrices $N_{i}$. 
For instance, we have
$$\sigma_{3} \otimes \sigma_{3} = \sigma_{3} + \sigma_{3} +  \sigma_{1} + 
\sigma_{1'} + \sigma_{1''}$$ and, at the same time, 
$$N_{3}.N_{3} = 2 N_{3} + N_{1} + N_{1'} + N_{1''}$$
In particular, the Dynkin diagram of $E_{6}^{(1)}$, considered as the 
fusion graph of the fundamental representation of the binary tetrahedral group (Mc 
Kay correspondence) can 
also be read in terms of the fusion matrices $N$:

$$
\begin{array}{ccc}
N_{2}.N_{1} = N_{2} &
N_{2}.N_{2} = N_{1} + N_{3} &
N_{2}.N_{3} = N_{2} + N_{2'} + N_{2''} \\
N_{2}.N_{2'} = N_{1'} + N_{3} &
N_{2}.N_{2''} = N_{1''} + N_{3} &
N_{2}.N_{1'} = N_{2'} \\
N_{2}.N_{1''} = N_{2''}
\end{array}
$$

\subsubsection{From the Grothendieck ring to the character table}

The seven commuting $7\times 7$ matrices $N_{i}$ can be simultaneously diagonalized with a
common similarity matrix $X$  (the rows of which give the 
eigenvalues of the  $N_i$).
$$
X = 
\left(
\begin{array}{ccccccc}
1& 1& 1& 1& 1& 1& 1\\ 
1& 1& 1& \omega^2& \omega^2& \omega& \omega\\ 
1& 1& 1& \omega& \omega& \omega^2& \omega^2\\ 
2& 0& -2& 1& -1& 1& -1\\ 
2& 0& -2& \omega^2& -\omega^2& \omega& -\omega\\ 
2& 0& -2& \omega& -\omega& \omega^2& -\omega^2\\ 
3& -1& 3& 0& 0& 0& 0
\end{array}
\right)
$$
A nice observation is that the character table is given by the 
same matrix $X$, \ie, by the list (properly ordered!) of eigenvalues
of the matrices $N_{i}$: each line corresponds to an irreducible 
representation and each column to a conjugacy class. The point is 
that we {\sl did not} have to work out the conjugacy classes themselves: the 
structure constants of the Grothendieck ring (which are themselves, in 
the present case, 
encoded by the graph $E_{6}^{(1)}$) provide enough data to reconstruct
the whole character table. This not so well-known result is true
 for any finite group.

\subsubsection{Perron-Frobenius data for the graph $E_{6}^{(1)}$}
We want now ``reverse'' the machine, \ie forget everything we know 
about groups $\tilde T$ (or $T$) and try to reconstruct as much as
we can from the combinatorial data provided by the $E_6^{(1)}$ diagram.
The adjacency matrix ${\cal G}$ of an oriented graph is a matrix labelled
by the vertices of ${\cal G}$ whose $(i,j)$ element is equal
to $n$ whenever there are $n$ edges from $i$ to $j$. 
When the graph is not oriented, each edge is considered as carrying both
orientations, so that the matrix of the graph is symmetric.

In our case, we label the vertices of the graph $E_6^{(1)}$ by $1,1',1'',2,2',2'',3$, in this order;
the corresponding adjacency matrix is clearly
$$
{\cal G} = 
\left( \begin{array}{ccccccc}
 0 & 0 & 0 & 1 & 0 & 0 & 0 \\
 0 & 0 & 0 & 0 & 1 & 0 & 0 \\
 0 & 0 & 0 & 0 & 0 & 1 & 0 \\
 1 & 0 & 0 & 0 & 0 & 0 & 1 \\
 0 & 1 & 0 & 0 & 0 & 0 & 1 \\
 0 & 0 & 1 & 0 & 0 & 0 & 1 \\
 0 & 0 & 0 & 1 & 1 & 1 & 0 
\end{array}
\right)
$$
This matrix is nothing else than the matrix $N_{2}$, \ie the 
fundamental generator of the ring of $N$-matrices and it is associated with the fundamental representaion $\sigma_2$.
The eigenvalues are  ${-2, -1, -1, 0, 1, 1, 2}$. The biggest eigenvalue (also called ''norm of the graph'' or
''Perron-Frobenius eigenvalue'') is equal to $2$. This is also true 
for the graphs $E_{7}^{(1)}$ and $E_{8}^{(1)}$.
The corresponding eigenvector (we normalize the first entry to $1$), also called ``Perron-Frobenius vector of the graph'' has
components $D$ that are positive integers.
$
D = {1, 1, 1, 2, 2, 2, 3}
.$
We recognize the dimensions of the irreps of $\tilde T$.

The conclusion is that, from the graph alone, we can recover the dimension 
of the irreps. This is also true for the binary cubic and icosahedral 
groups. From the same graph, we then recover, as already explained, the multiplication of representations
by the fundamental (which is $2$-dimensional) as well as
the whole fusion algebra, by imposing associativity of the Grothendieck ring;
the character table itsef is obtained by a simultaneous diagonalisation of the
$N_i$ matrices encoding the structure constants $C_{ijk}$ of this ring.

\subsubsection{Structure of centralizer algebras, tower of commutants for $\tilde T$}

Using the structure of the Grothendieck ring for $\tilde T$, encoded 
by the graph $E_6^{(1)}$, we see immediately, by taking tensor powers of the fundamental,  that
\begin{eqnarray*}
[2]^{1} & = & 1 [2] \cr
[2]^{2} & = & 1 [1] + 1 [3] \cr
[2]^{3} & = & 2 [2] + 1 [2'] + 1 [2''] \cr
[2]^{4} & = & 2 [1] + 4 [3] + 1 [1'] + 1 [1''] \cr
[2]^{5} & = & 6 [2] + 5 [2'] + 5 [2''] \cr
[2]^{6} & = & 6 [1] + 16 [3] + 5 [1'] + 5 [1''] \cr
\ldots & = & \etc
\end{eqnarray*}
We call $C_{p}$ the centralizer algebras of the group $\tilde T$ in the representation $[2]^{p}$.
It is clear that these algebras  are not isomorphic with the
usual Temperley-Lieb  algebras $T_{p}$ (which are isomorphic with the  Schur
centralizer algebras for $SU(2)$) as soon as $p \geq 3$. For instance $T_{4}
= M(2,\CC) \oplus M(3,\CC) \oplus \CC$ since, in $SU(2)$, $[2]^4 = 2[1] +
3[3] + 1 [5]$,  but $C_{4} = M(2,\CC) \oplus M(4,\CC) \oplus \CC \oplus \CC$
since,  in $\tilde T$, $[2]^{4} = 2 [1] + 4 [3] + 1 [1'] + 1 [1''] $.
The above results leads immediately to the following structure
(Fig. \ \ref{CommutantsT}) for the tower of commutants $C_{p}$'s. As usual, 
inclusions are defined by the edges of this graph and an appropriate Pascal rule 
gives the dimensions. Notice that after a few steps, we get the 
folded $E_6^{(1)}$ diagram of reflected and repeated 
down to infinity. This picture -- paths emanating from the endpoint 
vertex -- can also be generated very simply by
considering successive powers of the adjacency matrix $N_2$ of the 
Dynkin diagram $E_{6}^{(1)}$ acting on the (transpose) of
the vector $(1,0,0,0,0,0,0)$ characterizing the leftmost vertex 
(identity representation).

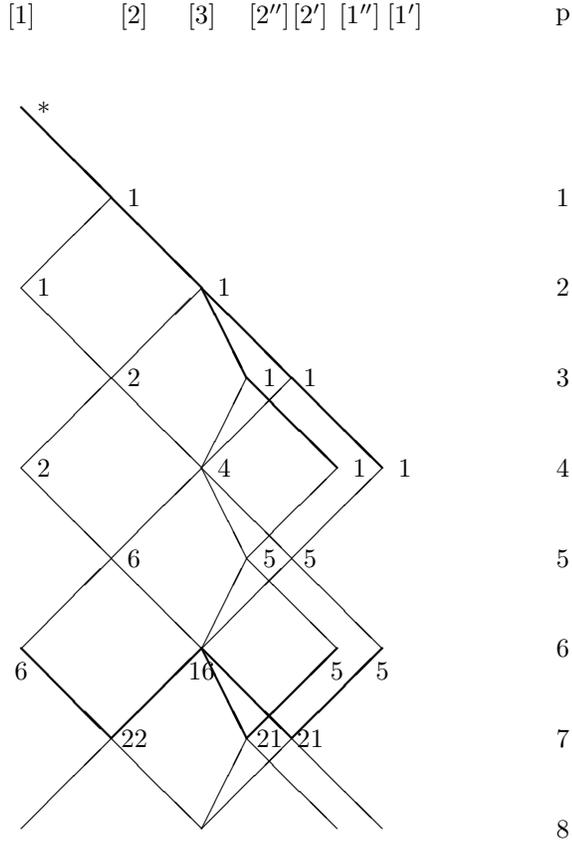
\vfill

\begin{figure}[htb]
\unitlength 0.6mm
\begin{center}
\begin{picture}(95,95)

\put(5,120){\makebox(0,0){$ \ast $}}
\thicklines
\put(0,120){\line(1,-1){20}}

\put(25,100){\makebox(0,0){$ 1 $}}
\thinlines
\put(0,80){\line(1,1){20}}
\thicklines
\put(20,100){\line(1,-1){20}}

\put(5,80){\makebox(0,0){$ 1 $}}
\put(45,80){\makebox(0,0){$ 1 $}}
\thinlines
\put(0, 80){\line(1,-1){20}}
\thicklines
\put(40,80){\line(1,-2){10}}
\put(40,80){\line(1,-1){20}}
\thinlines
\put(20,60){\line(1,1){20}}

\put(25,60){\makebox(0,0){$ 2 $}}
\put(65,60){\makebox(0,0){$ 1 $ }}
\put(55,60){\makebox(0,0){$ 1 $}}

\put(0,40){\line(1,1){20}}
\put(40,40){\line(1,1){20}}
\put(40,40){\line(1,2){10}}

\put(20,60){\line(1,-1){20}}
\thicklines
\put(50,60){\line(1,-1){20}}
\put(60,60){\line(1,-1){20}}
\thinlines
\put(5,40){\makebox(0,0){$ 2 $}}
\put(45,40){\makebox(0,0){$ 4 $}}
\put(75,40){\makebox(0,0){$ 1 $}}
\put(85,40){\makebox(0,0){$ 1 $}}

\put(0,40){\line(1,-1){20}}
\put(40,40){\line(1,-2){10}}
\put(40,40){\line(1,-1){20}}

\put(20,20){\line(1,1){20}}
\put(50,20){\line(1,1){20}}
\put(60,20){\line(1,1){20}}

\put(25,20){\makebox(0,0){$ 6 $}}
\put(65,20){\makebox(0,0){$ 5 $ }}
\put(55,20){\makebox(0,0){$ 5 $}}

\put(0,0){\line(1,1){20}}
\put(40,0){\line(1,1){20}}
\put(40,0){\line(1,2){10}}

\put(20,20){\line(1,-1){20}}
\put(50,20){\line(1,-1){20}}
\put(60,20){\line(1,-1){20}}

\put(0, - 5){\makebox(0,0){$ 6 $}}
\put(40, - 5){\makebox(0,0){$ 16 $}}
\put(70, - 5){\makebox(0,0){$ 5 $}}
\put(80,- 5){\makebox(0,0){$ 5 $}}

\thicklines

\put(0, 0){\line(1,-1){20}}
\put(40, 0){\line(1,-2){10}}
\put(40, 0){\line(1,-1){20}}

\put(20, -20){\line(1,1){20}}
\put(50,-20){\line(1,1){20}}
\put(60,-20){\line(1,1){20}}

\put(25,-20){\makebox(0,0){$ 22 $}}
\put(65,-20){\makebox(0,0){$ 21 $ }}
\put(55,-20){\makebox(0,0){$ 21 $}}

\thinlines

\put(0,-40){\line(1,1){20}}
\put(40,-40){\line(1,1){20}}
\put(40,-40){\line(1,2){10}}

\put(20,-20){\line(1,-1){20}}
\put(50,-20){\line(1,-1){20}}
\put(60,-20){\line(1,-1){20}}
\put(0, 140){\makebox(0,0){$ [1] $}}
\put(40, 140){\makebox(0,0){$ [3] $}}
\put(75, 140){\makebox(0,0){$ [1''] $}}
\put(85, 140){\makebox(0,0){$ [1'] $}}

\put(25,140){\makebox(0,0){$ [2] $}}
\put(65,140){\makebox(0,0){$ [2'] $ }}
\put(55,140){\makebox(0,0){$ [2''] $}}

\put(120,140){\makebox(0,0){p}}
\put(120,120){\makebox(0,0){}}
\put(120,100){\makebox(0,0){1}}
\put(120,80){\makebox(0,0){2}}
\put(120,60){\makebox(0,0){3}}
\put(120,40){\makebox(0,0){4}}
\put(120,20){\makebox(0,0){5}}
\put(120,0){\makebox(0,0){6}}
\put(120,-20){\makebox(0,0){7}}
\put(120,-40){\makebox(0,0){8}}
\end{picture}
\label{CommutantsT}
\end{center}
\vskip 2cm
\end{figure}


\subsubsection{Essential paths for the graph $E_6^{(1)}$}

We first define elementary paths on a graph as a sequence 
$\{\sigma_{a_1}, \sigma_{a_2}, \ldots \sigma_{a_p}\}$ of 
consecutive vertices (here, for simplicity, we suppose that edges carry both 
orientations, \ie no orientation at all). Elementary paths can 
therefore backtrack. Then we consider the Hilbert space $Paths$ 
obtained by taking arbitrary linear combinations of elementary paths.
The scalar product is defined by declaring that the basis of 
elementary paths is orthonormal. Since every elementary path has a 
length $p$, the vector space $Paths$ is graded\footnote{Warning:
the length of $[\sigma_a \sigma_b \sigma_c \sigma_b \sigma_c \sigma_d]$ is
$5$, not $3$.}. In the case of 
$SU(2)$ (with graph $A_{\infty}$) or $\tilde T$ (with graph $E_6^{(1)}$), 
irreducible representations appearing in the decomposition of $[2]^p$ can be 
characterized by  paths on those graphs, emanating from the 
origin; they are also associated with particular projectors, that are, 
$2^p \times 2^p$ matrices. We need now to introduce special paths called
{\sl essential} paths. The notion of essential paths on a graph is due to  Ocneanu (\cite{Ocneanu:paths}). Essential paths may start from any 
vertex but we shall be mostly interested in those starting from the 
origin.

Let us begin with the case of $SU(2)$.
A given irreducible representation of dimension $d$ appears for the 
first time in the decomposition of $[2]^{d-1}$ and corresponds
to a particular projector in the vector space $(\CC^2)^{\otimes d-1}$ 
which is totally symmetric and therefore projects on the
space of symmetric tensors. These symmetric tensors provide a basis 
of this particular representation space and are, of course, in
one to one correspondence with symmetric polynomials in two complex 
variables $u, v$ (representations of given degree).
Paths corresponding to irreducible symmetric representations are essential
paths starting at the origin.

However, irreducible representations of dimension $d$ appear not only in the reduction of 
$[2]^{d-1}$ but also in the reduction of $[2]^f$, when 
$f = d+1, d+3, \dots$. Such representations are equivalent with the 
symmetric representations previously described but they are
nevertheless distinct, as explicit given representations and 
their associated projectors are not symmetric.
For instance the representation $[3]$ that appears in $[2]^2$ 
corresponds to an essential path starting from the origin, but the
three representations $[3]$  that appear in the reduction of $[2]^4$ do not correspond to 
essential paths\footnote{
The notion of ``essential path'' on a graph $G$ formalizes and  generalizes
the above remarks. In the present paper, we shall only need to count these
particular paths, so that we do not need to give their precise definition.
The interested reader will find this information in the appendix.}.

When we move from the case of $SU(2)$ to the case of finite subgroups 
of $SU(2)$, in particular the binary polyhedral
groups whose representation theories are described by the affine Dynkin 
diagrams $E_6^{(1)}$, $E_7^{(1)}$ and $E_8^{(1)}$,
the notion of essential paths can be obtained very simply by 
declaring that a path on the corresponding diagram
is essential if it describes an irreducible representation that 
appears in the reduction with respect to the chosen 
finite subgroup of an irreducible {\sl symmetric} representation of $SU(2)$.
For instance, the $[4]$ dimensional representation of $SU(2)$, 
obtained in the decomposition of $[2] [3] = [2] + [4]$ is symmetric, 
and is associated with a Wenzl projector $p_{4}$ of the algebra $T_{4}$.
In the case of the finite subgroup $\tilde T$, the corresponding projector of the 
centralizer algebra $C_{4}$ splits, and this 
corresponds to the reduction $[4] \rightarrow [2'] + [2'']$ into a sum of two inequivalent irreps. 
In $SU(2)$ we have therefore one essential path of length $3$, emanating from the origin 
(it ends on $[4]$), but in $\tilde T$, this gives two essential paths, 
one ending on $[2']$ and the other on $[2'']$.
In general, essential paths are linear combinations of elementary 
paths.

The number of essential paths starting from the origin and
ending at a given vertex are readily obtained from the tower of centralizers 
by using a kind of ``moderated'' Pascal rule: the number of essential 
paths (with fixed origin) of length $p$ reaching a particular vertex
is obtained from the sum of the number of essential paths of length $p-1$ 
reaching the neighbouring points (as in Pascal rule) by substracting 
the number of paths of length $p-2$ reaching the chosen vertex.
This observation was made by \cite{Ocneanu:paths} in a general 
setting, and by J.B. Zuber (\cite{Zuber:Bariloche}) in the context of 
boundary conformal field theories.

The following picture -- essential paths starting
from the endpoint 
vertex  -- can be generated very simply as follows :
Define a rectangular matrix $E_{1}$, with seven columns and infinitely many rows, 
whose $j-th$ row is  $E_{1}(j) = N_2 E_{1}(j-1) - E_{1}(j-2)$, with 
$E_{1}(1) = (1,0,0,0,0,0,0)$, 
$E_{1}(2) = N_2 E_{1}(1)$, and where $N_2$ is the adjacency matrix of the graph $E_6^{(1)}$. 
The entries of $E_{1}(j)$ give the number of paths of length $j-1$ ending on the different vertices. 
Essential paths starting from
an arbitrary vertex of the Dynkin diagram can be constructed in the same way,
by replacing $E_{1}$ by $E_{2}(1) = (0,1,0,0,0,0,0)$,
  $E_{3}(1) = (0,0,1,0,0,0,0)$\etc.
Information about essential paths (in particular their number) is therefore 
encoded by the set of seven rectangular matrices (in the case of 
$E_{6}^{(1)}$) which have seven rows and infinitely many rows\footnote{In the
 following picture, we decided arbitrarily to cut the graph at level $p=8$}:
essential paths for the finite subgroups of $SU(2)$ can be of arbitrary length 
since symmetric representations of $SU(2)$ can
be of arbitrary degree (horizontal Young diagrams with an arbitrary 
number of boxes).  Their 
quantum analogues, however have only a finite number of rows. These 
matrices, introduced in \cite{Coquereaux:QuantumTetra}, will
 be called ``essential matrices

\vskip 4cm

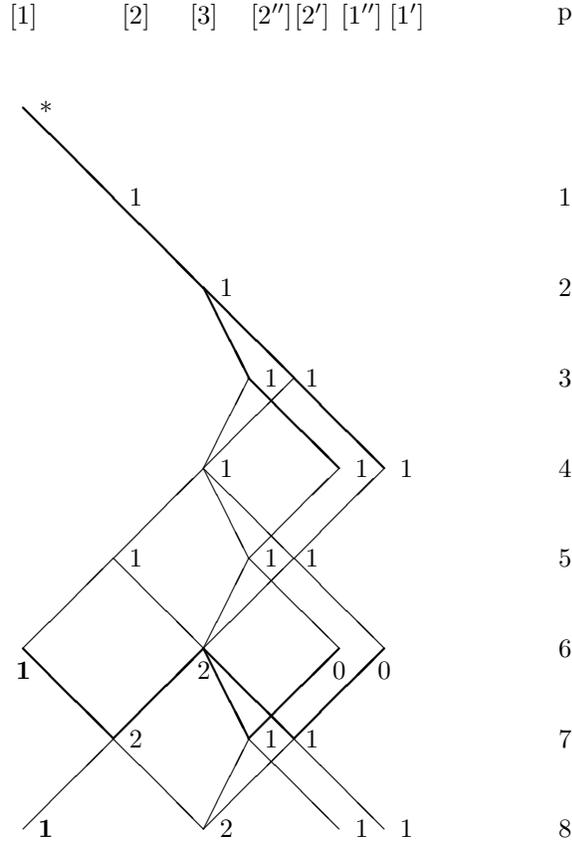
\begin{figure}[htb]
\unitlength 0.6mm
\begin{center}
\begin{picture}(87,87)

\put(5,120){\makebox(0,0){$ \ast $}}
\thicklines
\put(0,120){\line(1,-1){20}}

\put(25,100){\makebox(0,0){$ 1 $}}
\thinlines
\thicklines
\put(20,100){\line(1,-1){20}}

\put(45,80){\makebox(0,0){$ 1 $}}
\thinlines
\thicklines
\put(40,80){\line(1,-2){10}}
\put(40,80){\line(1,-1){20}}
\thinlines

\put(65,60){\makebox(0,0){$ 1 $ }}
\put(55,60){\makebox(0,0){$ 1 $}}

\put(40,40){\line(1,1){20}}
\put(40,40){\line(1,2){10}}

\thicklines
\put(50,60){\line(1,-1){20}}
\put(60,60){\line(1,-1){20}}
\thinlines
\put(45,40){\makebox(0,0){$ 1 $}}
\put(75,40){\makebox(0,0){$ 1 $}}
\put(85,40){\makebox(0,0){$ 1 $}}

\put(40,40){\line(1,-2){10}}
\put(40,40){\line(1,-1){20}}

\put(20,20){\line(1,1){20}}
\put(50,20){\line(1,1){20}}
\put(60,20){\line(1,1){20}}

\put(25,20){\makebox(0,0){$ 1 $}}
\put(65,20){\makebox(0,0){$ 1 $ }}
\put(55,20){\makebox(0,0){$ 1 $}}

\put(0,0){\line(1,1){20}}
\put(40,0){\line(1,1){20}}
\put(40,0){\line(1,2){10}}

\put(20,20){\line(1,-1){20}}
\put(50,20){\line(1,-1){20}}
\put(60,20){\line(1,-1){20}}

\put(0, - 5){\makebox(0,0){$ {\bf 1} $}}
\put(40, - 5){\makebox(0,0){$ 2 $}}
\put(70, - 5){\makebox(0,0){$ 0 $}}
\put(80,- 5){\makebox(0,0){$ 0 $}}

\thicklines

\put(0, 0){\line(1,-1){20}}
\put(40, 0){\line(1,-2){10}}
\put(40, 0){\line(1,-1){20}}

\put(20, -20){\line(1,1){20}}
\put(50,-20){\line(1,1){20}}
\put(60,-20){\line(1,1){20}}

\put(25,-20){\makebox(0,0){$ 2 $}}
\put(65,-20){\makebox(0,0){$ 1 $ }}
\put(55,-20){\makebox(0,0){$ 1 $}}

\thinlines

\put(0,-40){\line(1,1){20}}
\put(40,-40){\line(1,1){20}}
\put(40,-40){\line(1,2){10}}

\put(20,-20){\line(1,-1){20}}
\put(50,-20){\line(1,-1){20}}
\put(60,-20){\line(1,-1){20}}

\put(5, - 40){\makebox(0,0){$ {\bf 1} $}}
\put(45, - 40){\makebox(0,0){$ 2 $}}
\put(75, - 40){\makebox(0,0){$ 1 $}}
\put(85,- 40){\makebox(0,0){$ 1 $}}
\put(0, 140){\makebox(0,0){$ [1] $}}
\put(40, 140){\makebox(0,0){$ [3] $}}
\put(75, 140){\makebox(0,0){$ [1''] $}}
\put(85, 140){\makebox(0,0){$ [1'] $}}

\put(25,140){\makebox(0,0){$ [2] $}}
\put(65,140){\makebox(0,0){$ [2'] $ }}
\put(55,140){\makebox(0,0){$ [2''] $}}

\put(120,140){\makebox(0,0){p}}
\put(120,120){\makebox(0,0){}}
\put(120,100){\makebox(0,0){1}}
\put(120,80){\makebox(0,0){2}}
\put(120,60){\makebox(0,0){3}}
\put(120,40){\makebox(0,0){4}}
\put(120,20){\makebox(0,0){5}}
\put(120,0){\makebox(0,0){6}}
\put(120,-20){\makebox(0,0){7}}
\put(120,-40){\makebox(0,0){8}}

\end{picture}
\label{EssentialPathsT}
\end{center}
\vskip 2cm
\end{figure}

\subsubsection{Projectors on irreducible representations}
Our purpose, here, is to explain, in a nutshell, how to obtain explicitly a matrix expression for
the projectors $\varpi_p [s]$ mapping the reducible representation space $[2]^p$  to one of its irreducible 
subrepresentations $[s]$, for  $SU(2)$ or one of its finite subgroups. These are explicit $2^p\times 2^p$ matrices.
We do it for $SU(2)$ first. Here are the steps:
\begin{itemize}
\item Find an explicit matrix realization for the Jones' projectors $e_{i}$'s in the appropriate
Jones-Temperley-Lieb algebra $T_p$.

\item Express the minimal central projectors associated with the various blocks appearing in the  algebra $T_p$
in terms of the Temperley-Lieb' generators $e_{i}$. 
\item Call $A$ the classical antisymmetrizer of $SU(2)$ in the  
representation $[2]^2$ (it is a $4\times 4$ matrix). We obtain the projectors  $\varpi [s]$
from the minimal central projectors by replacing  $e_1, e_2,\ldots$'s by 
$\epsilon_1=  A \otimes \one_{2} \otimes \one_{2} \otimes \one_{2} \otimes \ldots$,
$\epsilon_2 = \one_{2} \otimes A \otimes \one_{2} \otimes \one_{2} \otimes \ldots$, \etc.
\end{itemize}
The sub-representation $[p+1]$ of $[2]^{p}$ is special since it 
is totally symmetric; the expression of the correspondig central projector of $T_p$
is easy to obtain in this case since a recurrence formula exists for Wenzl projectors (see for instance
\cite{Jones:book}).

Now, we do it for the binary tetrahedral group.
\begin{itemize}
	\item We first compute the projector decomposition of $[2]^p$
for $SU(2)$ as above.

	\item  We use the fact that the binary tetrahedral group is generated 
	by two explicit generators ${\bf s}$ and ${\bf t}$.  Considered as (group-like) 
	elements of the group algebra, \ie  $\Delta {\bf s} = {\bf s} \otimes {\bf s}$ and
        $\Delta {\bf t} =  {\bf t} \otimes  {\bf t}$, we can compute their iterated coproduct
in representation $[2]^p$,  they are explicit $2^p \times 2^p$ matrices. The 
	projectors  of $SU(2)$, that we have already obtained,  commute with them.
 To simplify the discussion, we take $p=3$ and write  $[2]^3= [2_a] + [2_b] + [4]$ for $SU(2)$, and $[4]\mapsto [2']+[2'']$ for the
reduction to $\tilde T$.
We want to obtain $\varpi_{3}[\sigma_{2'}]$  and $\varpi_{3}[\sigma_{2''}]$. 
	 Taking  an arbitrary $8 \times 8$ matrix $\varpi$, we 
	 first impose that it belongs 
	to the centralizer algebra, so it should commute with the iterated 
	coproduct of $ {\bf s}$ and $ {\bf t}$ already calculated (linear equations for 
	the matrix coefficients). This already restricts the number of 
	unknown coefficients.
	
	\item 
         We impose that the matrix $\varpi$ should be orthogonal to the known $\varpi_{3}[\sigma_{2a}]$ 
	and $\varpi_{3}[\sigma_{2b}]$ (again linear equations). This further restrict the number of 
	coefficients.

	\item  Finally we impose that $\varpi$ should be a projector
 ($\varpi.\varpi = \varpi$). 
	We find three solutions : one is of rank $4$ (it is 
	$\varpi_{3}[\sigma_{4}]$ itself), the other two are of rank $2$ and 
	add up to $\varpi_{3}[\sigma_{4}]$. These are the projectors 
	$\varpi_{3}[\sigma_{2'}]$  and $\varpi_{3}[\sigma_{2''}]$ that we 
	were looking for.
	
	  \end{itemize}
Writing even a simple example in full details requires a lot of room, but the procedure should be clear.

\subsubsection{Klein invariants}

Take a classical polyhedron, put its vertices $V$ on the sphere; from the centroid of the polyhedron draw
rays in direction of the points located at the center of the faces and at the middle of the edges. These
rays intersect the sphere at points $F$, and $E$. Notice (Euler) that $\# F-\# E+\# V=2$. Now
make a stereographical projection and 
build a complex polynomial that vanishes precisely at the location of the projected vertices (or center of faces, or mid-edges):
this polynomial is, by construction,  invariant under the symmetry group of the polyhedron (at least
projectively) since group elements only permute the roots.
 This is the historical method -- see in particular the famous little book ~\cite{Klein}.
In the case of the tetrahedron, for instance, you get the three polynomials (in homogeneous coordinates):
$V = u^4 + 2i\sqrt 3 u^2 v^2 + v^4$, $E = uv(u^4-v^4)$ and $F=u^4-2i\sqrt 3 u^2 v^2 + v^4$. Actually $V$ and $F$ are only
projectively invariant, but $X=108^{1/4} E $, $Y = - VF = -(u^8+v^8 + 14 u^{4} v^{4})$ and $Z = V^3 - i X^2
=(u^{12}+v^{12})-33(u^8 v^4 + u^4 v^8)$ are (absolute) invariants, of degrees $6,8,12$.
 Together with the relation $X^4+Y^3+Z^2=0$, they generate the whole set of invariants.
Alternatively you can build the $p$-th power of the fundamental representation  of the symmetry group
of the chosen binary polyhedral group, and choose $p$ such that there exists one  essential path of length $p$ starting at the origin
of the graph of tensorisation by the fundamental representation ( one of the affine $ADE$ diagrams) that returns to the origin.
Therefore you get a symmetric tensor (since the path is essential), hence a homogeneous polynomial of degree $p$; moreover this
polynomial is invariant since the path goes back to the origin (the identity representation). By explicitly
calculating the projectors corresponding to the (unique) essential path of $[2]^6$, $[2]^8$ and $[2]^{12}$ on the affine $E_6^{(1)}$ graph,
we can recover the polynomials $X,Y,Z$.

\subsection{Quantum geometry}

The main interest of the previous section was to show that a good deal 
of the geometry associated with symmetry groups of platonic solids could
be carried out without using the groups themselves but only the 
exceptionnal affine Dynkin diagrams.
Going to the quantum geometry will now be relatively easy: we just replace the
affine Dynkin diagrams by the usual Dynkin diagrams.
This present section will be rather  short. One reason is the limited amount
of space available for these proceedings, another reason is that the 
techniques have already been presented in the previous section, and 
they can be translated directly without further ado.
We shall mention only the differences with the previous 
(classical) situation.

\subsubsection{Realisations of the quantum algebras}

The quantum algebra analogues of the group algebras ${\cal H}_{E_6^{(1)}}$ and ${\cal H}_{E_8^{(1)}}$ should be
 objects called ${\cal H}_{E_{6}}$ and ${\cal H}_{E_{8}}$. 
The quantum analogue of ${\cal H}_{E_7^{(1)}}$ does not exist, for a reason 
explained later (but the
$E_{7}$ diagram leads nevertheless to an interesting quantum geometry).
Actually we shall not introduce such quantum algebras at all,
but we proceed as if they had been constructed. Besides the basic
reference \cite{Ocneanu:paths}, let us mention also the following 
papers: \cite{PetkovaZuber:Oc}, that uses the formalism of boundary 
conformal field theories, \cite{Evans}, for a discussion in terms of 
nets of subfactors and  \cite{KirillovJr}, for a discussion in terms of
braided modular categories. Let us finally mention the paper 
\cite{Jones:planar}, that introduces planar algebras, a concept
that probably allows one to accommodate many of these construcions.

\subsubsection{Representations of these quantum algebras}

Since we did not give any definition of these quantum tetrahedral or 
icosahedral algebras, we shall define their irreducible representations 
$\sigma_{p}$ as mere symbols associated with
vertices of the diagrams $E_{6}$ or $E_{8}$.

The norms of the adjacency matrix ${\cal G}$ of the three exceptionnal Dynkin 
diagram are no longer the integer $2$ but the quantum integers 
$[2]_{q}$ \ie $2 \cos(\pi/\kappa)$ for $\kappa = 12, 18,30$. $\kappa$ is also
the Coxeter number of the diagram\footnote{The quantum numbers are $[n] = \frac{q^n - q^{-n}}{q -
q^{-1}}$. For $E_6$, $q=exp(i \pi/12)$.}.

The components of the corresponding (normalized) eigenvector $D$
(Perron-Frobenius)
with the first entry normalized to $1$ (extremity of the longest leg), are not positive integers
but they provide a definition for the (quantum) dimensions of the 
irreducible representations. For $E_6$, $D = qdim(\sigma_0,\sigma_1,\sigma_2,\sigma_5,\sigma_4,\sigma_3)$ $=
([1],[2],[3],[2],[1],[3]/[2]=\sqrt 2)$.

To keep the size of this paper small enough and specify our 
conventions for the labelling of vertices, we only give the graphs
$E_{6}$ and $E_8$. Much more information concerning the ``quantum tetrahedron''
can be found in the paper \cite{Coquereaux:QuantumTetra}.
The diagrams $E_6$ and $E_8$, with our labelling for vertices (increasing
labels starting from the tips) are given as follows
\begin{figure}[htb]
\unitlength 0.6mm
\begin{picture}(95,35)
\thinlines 
\multiput(10,10)(15,0){5}{\circle*{2}}
\put(40,25){\circle*{2}}
\thicklines
\put(10,10){\line(1,0){60}}
\put(40,10){\line(0,1){15}}
\put(10,3){\makebox(0,0){$[\sigma_0]$}}
\put(25,3){\makebox(0,0){$[\sigma_1]$}}
\put(40,3){\makebox(0,0){$[\sigma_2]$}}
\put(55,3){\makebox(0,0){$[\sigma_5]$}}
\put(70,3){\makebox(0,0){$[\sigma_4]$}}
\put(48,27){\makebox(0,0){$[\sigma_3]$}}
\thinlines 
\multiput(100,10)(15,0){7}{\circle*{2}}
\put(160,25){\circle*{2}}
\thicklines
\put(100,10){\line(1,0){90}}
\put(160,10){\line(0,1){15}}
\put(100,3){\makebox(0,0){$[\sigma_0]$}}
\put(115,3){\makebox(0,0){$[\sigma_1]$}}
\put(130,3){\makebox(0,0){$[\sigma_2]$}}
\put(145,3){\makebox(0,0){$[\sigma_3]$}}
\put(160,3){\makebox(0,0){$[\sigma_4]$}}
\put(175,3){\makebox(0,0){$[\sigma_7]$}}
\put(190,3){\makebox(0,0){$[\sigma_6]$}}
\put(158,27){\makebox(0,0){$[\sigma_5]$}}
\end{picture}
\caption{The diagrams $E_6$ and ${E_8}$}
\label{graphE6}
\end{figure}
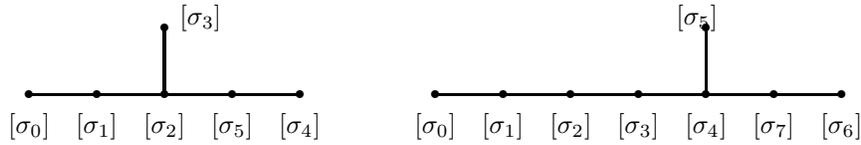

\subsubsection{Tensor products of representations and structure of the Grothendieck ring }

The representation $\sigma_{0}$ associated with the end-point of the longest leg is the unit.
The Dynkin diagram defines multiplication by the algebraic generator $\sigma_{1}$.
The only task is to complete the table (which is $6\times 6$
 in the case of the $E_{6}$ diagram) as we did in 
the classical case. This is rather straightforward and works perfectly, but 
for the fact that the obtained structure constants are positive integers for 
$E_{6}$ and $E_{8}$, but not for $E_{7}$. 
This was observed long ago by \cite{Pasquier}.
 Here is the ``fusion table''  we get 
for $E_{6}$:
	
$$
\begin{array}{||c||c|c|c|c|c|c||}
\hline
{} & 0 & 3  & 4 & 1 & 2 & 5   \\
\hline
\hline
0 & 0 & 3 & 4  & 1 & 2 & 5 \\
3 & 3 & 04  & 3 & 2 & 1 5 & 2 \\
4 & 4 & 3 & 0 & 5 & 2 & 1 \\
1 & 1 & 2 & 5 & 0 2 & 1 3 5 & 2 4   \\
2 & 2 & 1 5 & 2 & 1 3 5 & 0224 & 1 3 5 \\
5& 5  & 2 & 1 & 2 4 & 1 3 5 & 0 2 \\
\hline
\end{array}
$$
  
\subsubsection{The fusion matrices $N_{i}$}

These six $6\times 6$ matrices can be worked out easily, as we did 
in the case of the binary tetrahedral group.
They can be found in \cite{Coquereaux:QuantumTetra}.

\subsubsection{From the Grothendieck ring to the character table}

Again, there is no drastic difference with the classical situation 
and we get a quantum character table\ldots
but for the fact that conjugacy classes are not even defined!
This defines a kind of Fourier transform which is both finite and quantum;
 its relation with the action of the modular group should be further 
investigated.

\subsubsection{Structure of centralizer algebras and tower of commutants 
for ${\cal H}_{E_{6}}$}

We can take the powers of $\sigma_{1}$ and decompose them into 
irreducibles, as we did in the classical case, just by using the 
adjacency matrix and the above fusion table. The coefficients -- multiplicities -- define the 
would-be centralizer algebras; they could be defined as multi-towers 
algebras associated with the chosen Dynkin diagram thought of as a 
Bratelli diagram, see \cite{Jones:book}. The next picture displays the 
first eight rows (one can go down to infinity).

\begin{figure}[htb]
\unitlength 0.6mm
\begin{center}
\begin{picture}(95,95)

\put(5,120){\makebox(0,0){$ \ast $}}
\thicklines
\put(0,120){\line(1,-1){20}}

\put(25,100){\makebox(0,0){$ 1 $}}
\thinlines
\put(0,80){\line(1,1){20}}
\thicklines
\put(20,100){\line(1,-1){20}}

\put(5,80){\makebox(0,0){$ 1 $}}
\put(45,80){\makebox(0,0){$ 1 $}}
\thinlines
\put(0, 80){\line(1,-1){20}}
\thicklines
\put(40,80){\line(1,-2){10}}
\put(40,80){\line(1,-1){20}}
\thinlines
\put(20,60){\line(1,1){20}}

\put(25,60){\makebox(0,0){$ 2 $}}
\put(65,60){\makebox(0,0){$ 1 $ }}
\put(55,60){\makebox(0,0){$ 1 $}}

\put(0,40){\line(1,1){20}}
\put(40,40){\line(1,1){20}}
\put(40,40){\line(1,2){10}}

\put(20,60){\line(1,-1){20}}
\thicklines
\put(60,60){\line(1,-1){20}}
\thinlines
\put(5,40){\makebox(0,0){$ 2 $}}
\put(45,40){\makebox(0,0){$ 4 $}}
\put(85,40){\makebox(0,0){$ 1 $}}

\put(0,40){\line(1,-1){20}}
\put(40,40){\line(1,-2){10}}
\put(40,40){\line(1,-1){20}}

\put(20,20){\line(1,1){20}}
\put(60,20){\line(1,1){20}}

\put(25,20){\makebox(0,0){$ 6 $}}
\put(55,20){\makebox(0,0){$ 4 $}}
\put(65,20){\makebox(0,0){$ 5 $ }}

\put(0,0){\line(1,1){20}}
\put(40,0){\line(1,1){20}}
\put(40,0){\line(1,2){10}}

\put(20,20){\line(1,-1){20}}
\put(60,20){\line(1,-1){20}}

\put(0, - 5){\makebox(0,0){$ 6 $}}
\put(40, - 5){\makebox(0,0){$ 15 $}}
\put(80,- 5){\makebox(0,0){$ 5 $}}

\thicklines

\put(0, 0){\line(1,-1){20}}
\put(40, 0){\line(1,-2){10}}
\put(40, 0){\line(1,-1){20}}

\put(20, -20){\line(1,1){20}}
\put(60,-20){\line(1,1){20}}

\put(25,-20){\makebox(0,0){$ 21 $}}
\put(55,-20){\makebox(0,0){$ 15 $}}
\put(65,-20){\makebox(0,0){$ 20 $ }}

\thinlines

\put(0,-40){\line(1,1){20}}
\put(40,-40){\line(1,1){20}}
\put(40,-40){\line(1,2){10}}

\put(20,-20){\line(1,-1){20}}
\put(60,-20){\line(1,-1){20}}

\put(0, - 35){\makebox(0,0){$ 21 $}}
\put(40, - 35){\makebox(0,0){$ 56 $}}
\put(80,- 35){\makebox(0,0){$ 20 $}}

\put(0, 140){\makebox(0,0){$ \tau_0 $}}
\put(25,140){\makebox(0,0){$ \tau_1 $}}
\put(40, 140){\makebox(0,0){$ \tau_2$}}
\put(55,140){\makebox(0,0){$\tau_3 $}}
\put(65,140){\makebox(0,0){$ \tau_5 $ }}
\put(75, 140){\makebox(0,0){$ \tau_4 $}}
\put(85, 140){\makebox(0,0){$ $}}

\put(120,140){\makebox(0,0){n}}
\put(120,120){\makebox(0,0){}}
\put(120,100){\makebox(0,0){1}}
\put(120,80){\makebox(0,0){2}}
\put(120,60){\makebox(0,0){3}}
\put(120,40){\makebox(0,0){4}}
\put(120,20){\makebox(0,0){5}}
\put(120,0){\makebox(0,0){6}}
\put(120,-20){\makebox(0,0){7}}
\put(120,-40){\makebox(0,0){8}}
\end{picture}
\label{Commutants}
\end{center}
\vskip 1.5cm
\end{figure}


\subsubsection{Essential paths}

Here we need to know the precise general  definition (Ocneanu) of essential paths if 
we want to find them explicitly (see the appendix or \cite{Ocneanu:paths}), 
but if we want only to count them, 
the method explained in the classical case works. The only difference 
with the classical case is that essential matrices, which are 
rectangular, with infinitely many rows in the classical case, 
are defined as matrices with only $\kappa -1$ rows in the quantum 
situation (if the matrices were not truncated at that level, 
the line $\kappa$ would be filled with $0$ and the
next ones would contain negative integers).  This reflects the fact 
that essential paths having a bigger length do not exist. For the $E_6$ diagram, $\kappa = 12$.

The six rectangular matrices $E_{a}$ of size $11\times 6$ describing essential 
paths on $E_{6}$ are explicitly given in \cite{Coquereaux:QuantumTetra}.
Columns of the essential matrices are labelled by the length $p$
of the paths, so $p$ runs from $0$ to $10$. This can be seen as a kind 
of labelling by the vertices of the Dynkin diagram $A_{11}$. Essential 
matrices have therefore columns labelled by $E_{6}$ and rows labelled by 
$A_{11}$. This is actually more than a simple remark since the 
algebra generated by the eleven $6\times 6$ square matrices $F_{i}(a,b) \doteq E_{a}(i,b)$ provides
 a representation of the fusion algebra of the graph $A_{11}$.
A pictorial description of essential paths for all $ADE$ Dynkin diagrams
can be found in the appendix of \cite{Ocneanu:paths}.
The essential matrix describing (essential) paths leaving the origin 
leads to the next  picture.


\begin{figure}[htb]
\unitlength 0.6mm
\begin{center}

\begin{picture}(90,90)

\thicklines

\put(5,120){\makebox(0,0){$ 1 $}}

\put(0,120){\line(1,-1){20}}

\put(25,100){\makebox(0,0){$ 1 $}}


\put(20,100){\line(1,-1){20}}

\put(45,80){\makebox(0,0){$ 1 $}}


\put(40,80){\line(1,-2){10}}
\put(40,80){\line(1,-1){20}}


\put(65,60){\makebox(0,0){$ 1 $ }}
\put(55,60){\makebox(0,0){$ 1 $}}

\put(40,40){\line(1,1){20}}
\put(40,40){\line(1,2){10}}


\put(60,60){\line(1,-1){20}}

\put(45,40){\makebox(0,0){$ 1 $}}
\put(85,40){\makebox(0,0){$ 1 $}}

\put(40,40){\line(1,-1){20}}

\put(20,20){\line(1,1){20}}
\put(60,20){\line(1,1){20}}

\put(25,20){\makebox(0,0){$ 1 $}}
\put(65,20){\makebox(0,0){$ 1 $ }}

\put(0,0){\line(1,1){20}}
\put(40,0){\line(1,1){20}}

\put(20,20){\line(1,-1){20}}

\put(0, - 5){\makebox(0,0){$ 1 $}}
\put(40, - 5){\makebox(0,0){$ 1 $}}

\put(0, 0){\line(1,-1){20}}
\put(40, 0){\line(1,-2){10}}

\put(20, -20){\line(1,1){20}}

\put(25,-20){\makebox(0,0){$ 1 $}}
\put(55,-20){\makebox(0,0){$ 1 $}}

\put(40,-40){\line(1,2){10}}

\put(20,-20){\line(1,-1){20}}

\put(40, - 35){\makebox(0,0){$ 1 $}}

\put(40, -40){\line(1,-1){20}}
\put(60, -60){\line(1,-1){20}}

\put(55,-60){\makebox(0,0){$ 1 $}}
\put(85, -80){\makebox(0,0){$ 1 $}}

\put(0, 140){\makebox(0,0){$ \tau_0 $}}
\put(25,140){\makebox(0,0){$ \tau_1 $}}
\put(40, 140){\makebox(0,0){$ \tau_2$}}
\put(55,140){\makebox(0,0){$\tau_3 $}}
\put(65,140){\makebox(0,0){$ \tau_5 $ }}
\put(75, 140){\makebox(0,0){$ \tau_4 $}}
\put(85, 140){\makebox(0,0){$ $}}

\put(120,140){\makebox(0,0){n}}
\put(120,120){\makebox(0,0){0}}
\put(120,100){\makebox(0,0){1}}
\put(120,80){\makebox(0,0){2}}
\put(120,60){\makebox(0,0){3}}
\put(120,40){\makebox(0,0){4}}
\put(120,20){\makebox(0,0){5}}
\put(120,0){\makebox(0,0){6}}
\put(120,-20){\makebox(0,0){7}}
\put(120,-40){\makebox(0,0){8}}
\put(120,-60){\makebox(0,0){9}}
\put(120,-80){\makebox(0,0){10}}
\end{picture}

\label{EssentialPathsGraphE0}
\end{center}
\vskip 5. cm
\end{figure}


\subsubsection{Projectors on irreducible representations}

The method is essentially the same as in the classical situation, but for the fact that 
symmetrizer and antisymmetizer of $SU(2)$ have to be replaced by
their quantum analogues, which can be obtained from the spectral
decomposition of the flipped $R$-matrix of the quantum $SU(2)$
Hopf algebra (the one that obeys the braid group relation). However,
when $p$ increases (when we reach the triple point) one needs to know how to
split the Wenzl projector. This was done, in the classical situation, 
by  explicitly using the generators  of the binary polyhedral groups 
and the expression of their iterated coproducts.
Here, the problem is still open, since we do not
have an explicit realization for the quantum algebras 
${\cal H}$.

\subsubsection{Klein invariants}

In the quantum situation, Klein invariants are {\sl defined} 
(\cite{Ocneanu:paths}) as essential paths starting from the origin and coming 
back to the origin. From the first essential matrix of $E_{6}$ or 
from the above equivalent corresponding graph of essential paths, we see that such 
an invariant exist (its length $n$ is equal to $6$). We can 
actually compute it explicitly as a path, \ie as a linear 
combination of elementary paths on the graph $E_{6}$. It would be 
nice to exhibit also a kind of homogeneous (but non-commutative) polynomial 
that implements it. This has not be obtained so far.

\section{Classical and quantum induction-restriction}

The purpose of these two next subsections
 is to investigate induction-restriction
theory of representations in two classical situations (one finite,
and the other infinite dimensional); namely we take the group
$\tilde T$ as a subgroup of $SU(2)$ or as a subgroup of $\tilde I$.
As we know, representation theory of these three classical objects
is encoded by the  Dynkin diagrams $E_6^{(1)}$, $E_8^{(1)}$ and $A_\infty$.

The purpose of the last subsection is to investigate a similar situation
in a quantum (but finite) case, encoded by the two graphs $E_6$ and $A_{11}$.

\subsection{Classical induction-restriction: $E_8^{(1)}$ versus $E_6^{(1)}$, 
\ie $\tilde I$ versus $\tilde T$}

\subsubsection{Classical branching rules $\tilde I \rightarrow \tilde T$}

Both groups are finite subgroups of $SU(2)$ and we have inclusions: 
$\tilde I \subset \tilde T \subset SU(2)$.
In both cases (diagrams $E_8^{(1)}$ and $E_6^{(1)}$) we have given the dimensions of the corresponding 
representations.
Representations of $SU(2)$ can be restricted to its subgroups, and, 
in the same way, irreducible representations of
$\tilde I$ can be restricted to $\tilde T$, and decomposed into 
irreducible representations of the latter.
The following table gives the branching rules $\tilde I   \rightarrow    \tilde T $:
$$
\begin{array}{ccccccccc}
1 &  \rightarrow & 1 ; & 2 &  \rightarrow & 2 ; & 3 &  \rightarrow & 3 \\
4 &  \rightarrow & 2' + 2'' ; & 5 &  \rightarrow  & 3 + 1 + 1'' ; &
6 &  \rightarrow & 2 + 2' + 2'' \\
4' &  \rightarrow & 3 + 1 ; & 3' &  \rightarrow & 3 ; & 2' &  \rightarrow & 2 \\
\end{array}
$$
It is easy to obtain the above table: one should just compare
 multiplications of  irreps of $SU(2)$,  $\tilde I$,
 or $\tilde T$ by the $2$-dimensional fundamental representation. 
For instance,
$2\times 3 = 2 + 4$ in $SU(2)$, but $2+2'+2''$ in $\tilde T$ and $2 + 
4$ in $\tilde I$; therefore, with respect to
the branching $\tilde I \rightarrow \tilde T$ we have $4 \rightarrow 
2'+2''$. 
Then we have $2 \times 4 = 3 + 5$ in $SU(2)$, but $2 \times (2' + 
2'') = 1 + 3 + 1' + 3 + 1'' + 3$ in $\tilde T$, whereas $2 \times 4 = 
3 + 5$ in $\tilde I$; therefore, we get $5 \rightarrow 3+1'+1''$ for the
 branching $\tilde I \rightarrow \tilde T$. In the same way we get 
$2\times 5 = 4 + 6$ both in $SU(2)$, and
in $\tilde I$, the restriction to $\tilde T$ reads $2\times 
(3+1'+1'')=2+2'+2''+2'+2''$, but we already know that
$4 \rightarrow 2'+2''$ therefore, we find $6 \rightarrow 2+2'+2''$.  
Next we have $2\times 6 = 5+4'+3'$ in $\tilde I$,
so that restriction of both sides to $\tilde T$ gives $1+3+1'+3+1''+3 
= 3+1'+1'' + (4'+3')_{\tilde T}$ and we find
 $4'+3' \rightarrow 3 + 3 +1$, so that the only possibility is 
$4'\rightarrow 3+1$ and $3' \rightarrow 3$.
The last branching rules can be obtained by restricting $2\times 4' = 
6+2'$ to $\tilde T$: we get $2\times (3+1) = 2+2'+2'' + 
(2')_{\tilde T}$ \ie $2 + 2' + 2'' + 2 = 2+2'+2'' + (2')_{\tilde T}$ 
and therefore the restriction $2' \rightarrow 2$.

\subsubsection{Sections of classical vector bundles over ${\tilde I}/{\tilde T}$}

Since $\tilde T$ is a subgroup of $\tilde I$, we can write the binary 
icosahedral group as a principal bundle over
the quotient  ${\tilde I}/{\tilde T}$, with structure group 
(typical fiber) $\tilde T$, 
the binary tetrahedral group.
For each representation of the structure group, in particular for 
each irreducible representation $\rho$ (and
carrier space $V_\rho$) of
 $\tilde T$ (we know that there are $7$ of them), we can build an 
associated vector bundle
$\tilde I \times_{\tilde T} V_\rho$, with basis  ${\tilde 
I}/{\tilde T}$ and typical fiber the vector
space $V_\rho$. Now we may consider the spaces of sections 
$\Gamma_\rho$ of those bundles, which are
functions on the finite homogenous space  ${\tilde I}/{\tilde T}$ 
and valued in the corresponding vector spaces.
In the particular case of the trivial representation of $\tilde T$ 
(called $1$), the carrier vector space is $\CC$
and the sections of $\Gamma_1$ just coincide with the space of 
complex valued functions on the finite
set  ${\tilde I}/{\tilde T}$, whose cardinality is $5=120/24$.

Here we are in a finite dimensional situation, but  Peter Weyl 
theory of induced representations still applies.
Let us take the example of $\Gamma_3$, the space of sections of $ 
\tilde I \times_{\tilde T} V_3$; this vector bundle
is a collection of five vector spaces of dimension $3$ (one 
above each of the five points of the coset);
the dimension of $\Gamma_3$ is therefore $3\times 5 = 15$. This 
fifteen dimensional space is the carrier space of
a natural representation of $\tilde I$ (the one induced by this 
particular vector bundle), but this representation
is not irreducible: its decomposition, in irreps of $\tilde I$, can 
be obtained from the previously given
table of branching rules:  $3$ (of $\tilde T$) appears on the right-hand 
side of the branching rules corresponding
to the irreps $3,5,4'$ and $3'$ of $\tilde I$, from this information, 
one deduces that $\Gamma_3 = [3]\oplus [5]\oplus [4']\oplus [3']$ (whose
sum is indeed $5\times 3 = 15$, as it should).
This induction process leads to the following table, 
for the decomposition of the various
spaces of sections (the vector spaces $\Gamma_\rho$) in irreducible 
representations of $\tilde I$:
$$
\begin{array}{|c|c|c|}
    \hline
\rho &  dim(\Gamma_\rho)&  \Gamma_\rho \\
  \hline
1  & 5 \times 1 = 5 & 1 + 4' \\
2  & 5 \times 2 = 10 & 2 + 6 + 2' \\
3  & 5 \times 3 = 15 & 3 + 5 + 4' + 3' \\
2'  & 5 \times 2 = 10 & 4 + 6 \\
1'  & 5 \times 1 = 5 & 5  \\
2'' & 5 \times 2 = 10 & 4 + 6 \\
1''  & 5 \times 1 = 5 & 5  \\
  \hline
\end{array}
$$
In particular, the space of 
functions on the finite set (five points) 
$\tilde I / \tilde T$, that we may call $Fun(\tilde I / \tilde T) \equiv 
\Gamma_1$ decomposes into
irreps of $\tilde I$ as  $4'+1$.

 In our case (the space of sections of vector bundles above
the finite {\sl left} homogeneous space $\tilde I /\tilde T$) we see that 
$\sum dim(\Gamma_p) = 5 + 10 + 15 + 10 +5 + 10 +5 = 60$, which is one 
half of the order of the group $\tilde I$ (by considering both left 
and right bundles,
we would get $120 = \# \tilde I$). 

 For us, the main interest of the previous 
remarks, is that, knowing only the dimensions
of the spaces of sections, we could
recover the order of $\tilde I$
and the order of $\tilde T$.
The order of $\tilde I$, namely $120$, is obtained by 
by taking the double of the sum of the dimensions of the spaces of sections.
The cardinality of the quotient  (namely $dim \Gamma_1=5$) is then obtained
 by summing the dimensions 
appearing in the decomposition of the space
of sections associated with the trivial representation.
The order of $\tilde T$, namely $24$, is finally obtained by dividing the order
of $\tilde I$ by the
the cardinality of the quotient.

Once  $dim \Gamma_1=5$ is known, we can recover
the dimensions of the irreducible representations ${\rho}$
themselves by taking the 
ratio $dim(\Gamma_{\rho}) / dim(\Gamma_{1})$.

\subsection{Classical induction-restriction: $A_{\infty}$ versus $E_{6}^{(1)}$ (\ie $SU(2)$ versus $\tilde T$)}

We restrict the representations of $SU(2)$
to irreps of $\tilde T$. 
We build the principal bundle $SU(2)$ as a 
$\tilde T$ bundle over the quotient $SU(2) / \tilde T$ (which is a 
three dimensional manifold) and
consider the ( seven) associated vector bundles relative to the 
seven irreps of $\tilde T$. We have therefore one such
 vector bundle for every point of the extended Dynkin diagram 
$E_6^{(1)}$. The only difficulty is to compute the branching rules.
One method is to proceed step by step, \ie
to use the information provided by the two Dynkin diagrams encoding 
tensor multiplication by the fundamental representation, computing tensor products
of irreps both for $SU(2)$ and its finite subgroup and comparing the results.
The easiest method  is to use essential matrices (they 
have infinitely many rows, in the present case); 
another technique -- which amounts to the same but is aesthetically 
more appealing --  is to draw the essential 
paths on $E_6^{(1)}$.
The only relevant essential matrix 
(for our present purpose) is the one labelled by the trivial  representation of $\tilde T$
\ie by the space of essential paths emanating from the leftmost point of the Dynkin diagram $E_6^{(1)}$.
 For instance the line $p=8$ of that graph (referring to $[2]^8$ \ie 
to the representation of dimension $9$ of $SU(2)$, 
and associated with an essential path 
of length $8$ on the graph $A_{\infty}$) tells us that $[9] 
\rightarrow [1] + 2 [3] + [1'] + [1'']$ in the branching $SU(2)$ versus $\tilde T$.
In other words, in order to perform reduction, we read the first 
essential matrix ``horizontally'', \ie we look at representations of 
$SU(2)$ given by  {\sl symmetric} polynomials of degree $n$ 
and branch them to this particular subgroup. In order to perform 
induction, we look at the same essential matrix, but ``vertically'', \ie 
we choose a particular irrep of $\tilde T$ (a particular column) and 
see for which values  it appears in branching rules of $[p+1]$. 

\subsubsection{Sections of vector bundles over $SU(2)/\tilde T$}
Call $V_\rho$ the vector space carrying the irreducible representation
$\rho$ of $\tilde T$  and
 $\Gamma_\rho$ the space of sections of the
homogeneous vector bundle $SU(2) \times_{\tilde T} V_{\rho}$.
The spaces of sections of these vector bundles 
can be decomposed as follows into irreducible representations
of $SU(2)$  (subscript give the multiplicity):
\begin{eqnarray*}
  \Gamma_1  & = & 1 + 7 + 9 + 13_2 + 15 + 17 +\ldots = Fun({SU(2)/ \tilde T})\\
  \Gamma_2  & = & 2 + 6 + 8_2 + 10 + 12_2 + 14_3 + 16_2 + \ldots \\
  \Gamma_3 & = & 3+5+7_2+9_2 + 11_3+13_3 +15_4 + 17 +\ldots \\
  \Gamma_2'  & = &4+6+8+10_2+12_2+14_2+16_3 +\ldots \\
  \Gamma_1'  & = &5+9+11+13+15+17_2 +\ldots \\
  \Gamma_2''  & = &4+6+8+10_2+12_2+14_2+16_3 +\ldots \\
  \Gamma_1''  & = &5+9+11+13+15+17_2 +\ldots 
\end{eqnarray*}

The degree of the homogenous polynomials providing a basis for a 
representation space of dimension $d$ is $d - 1$, so that representations of degree 
$0,6,8,12, 14,16, \ldots$ appear in $\Gamma_1$, as it should: we 
recover the fact that these representations of  $SU(2)$ indeed contain
$\tilde T$ -  invariant subspaces (Klein polynomials for the tetrahedron).

We can make the same kind of comments as in the previous section
for instance $dim(\Gamma_3)/dim(\Gamma_1)=3$, 
 but we now have to manoeuvre infinite sums and we should use
generating functions (we shall not do it here).
In the case of $\Gamma_3$, for instance, we can also write
$$[3] \otimes ([1] \oplus [7] \oplus [9] \oplus 2 [13] \oplus 
\ldots )=[3] \oplus [5] \oplus 2[7] \oplus \ldots $$

The reader may wonder why we did not introduce also ``essential 
matrices'' in the previous subsection (with columns labelled by
irreps of the binary tetrahedral group and rows  labelled by irreps 
of the binary icosahedral group). There is no reason: we could have done 
it as well.

\subsection{Quantum induction-restriction:  $A_{11}$ versus $E_6$}

We now  replace the diagram $A_\infty$ that 
describe irreps of $SU(2)$ by the diagram $A_{11}$
and the classical binary tetrahedral group by its would-be 
quantum counterpart described by the diagram $E_6$.

Both examples studied in the corresponding classical two  sections actually provide interesting
-- and complementary -- classical analogues: the first ($ E_{8}^{(1)} \rightarrow 
E_{6}^{(1)}$) because it is finite dimensional, and the other 
$ A_{\infty} \rightarrow E_{6}^{(1)}$ because $A_{11}$ looks 
indeed as a ``truncated'' $A_\infty$.

From the embedding $\tilde T \subset \tilde I$, 
we can deduce  an embedding of the 
corresponding group algebras (finite dimensional Hopf algebras)
${\cal H}_{\tilde T} = \CC \tilde T \subset  {\cal H}_{\tilde I}  = \CC \tilde I$
but although we do not plan to give a construction here of the ``would-be groups'' 
(or would-be group algebras)
that we could associate with the two genuine Dynkin diagrams $E_6$ 
and $A_{11}$,
 we want nevertheless to consider the first as a kind
of sub-object of the next. We proceed as if we had an embedding ${\cal H}_{E_6}  \subset  {\cal H}_{A_{11}} .$

We take $\hat q = exp (i \pi/12)$ (so that if $q = {\hat q}^2$, 
then $q^{12}=1$).
Irreps of $A_{11}$ are representations called $\tau_0, \tau_1, 
\ldots \tau_{10}$. They have $q--$dimension respectively
 equal to  $[1],[2],[3],[4],[5],[6],[7]=[5],[8]=[4],[9]=[3],[10]=[2],[11]=[1]$.
There actually is a non semi-simple Hopf algebra defined as a finite
dimensional quotient of the enveloping quantum algebra of $SU(2)$, 
when $q$ is a twelfth root of unity, and which is such that the above 
list of $\tau_{i}$ indeed labels its irreducible representations of non-zero quantum dimension;
however this knowledge will not be used here. The ``representations'' $\tau_{i}$ are 
therefore just abstract symbols that we can associate with the various points of 
the diagram $A_{11}$, whose own fusion table could have been worked out
as discussed previously, and the ``$q$-dimensions'' is just a name for the entries of 
the normalized eigenvector associated with the norm of the adjacency 
matrix of this diagram (the norm being, by definition, its biggest 
eigenvalue).
Remember that both graph $A_{11}$ and $E_6$ have same norm.
 A priori, the ring  of representations that we are 
considering here have a ``dimension function''  valued in  a $\ZZ$-ring
linearly generated by the $q$-integers $[1],[2],[3],[4],[5],[6].$
This is a clearly the case both for $A_{11}$ and $E_6$.

It may be useful to note that 
$$[1]=1,[2]=\frac{\sqrt 2}{\sqrt 3 -1},[3]=\frac{ 
2}{\sqrt 3 -1},[4]=\frac{\sqrt 6}{\sqrt 3 -1},[5]=\frac{1+ \sqrt 
3}{\sqrt 3 -1},[6]=\frac{2 \sqrt 2}{\sqrt 3 -1}$$

\subsubsection{Quantum branching rules $A_{11} \rightarrow E_6$}

The branching rules from $A_{11}$ to $E_6$ (that gave restriction in 
one direction and induction in the other) are
 gotten from the ``$q$-symmetric'' representations,
or, equivalently, from the essential matrices of the graph $E_6$. 

The only relevant essential matrix, for our present purpose, is the one 
labelled by the ``trivial representation'' (leftmost point of the graph 
$E_{6}$).
The following table summarizes the results for the reduction
$A_{11}  \rightarrow  E_6$. One should remember that  the 
$q$-dimension corresponding to irreps $\tau_p$ of $A_{11}$, (\ie vertices of $A_{11}$)
is $[p+1]$.
$$
\begin{array}{|ccc|ccc|ccc|}
\tau_0   &  \rightarrow & \sigma_0 & 
\tau_1  &  \rightarrow &  \sigma_1 & 
\tau_2  &  \rightarrow &  \sigma_2 \\

\tau_3  &  \rightarrow &  \sigma_3 + \sigma_5 &
\tau_4  &  \rightarrow &  \sigma_2 + \sigma_4 &
{} & {} & {} \\

\tau_5  &  \rightarrow &  \sigma_1 + \sigma_5 &
{} & {} & {} &
{} & {} & {} \\

\tau_6  &  \rightarrow &  \sigma_0 + \sigma_2 &
\tau_7  &  \rightarrow &  \sigma_1 + \sigma_3 &
{} & {} & {} \\

\tau_8  &  \rightarrow &  \sigma_2 &
\tau_9  &  \rightarrow &  \sigma_5 &
\tau_{10}  &  \rightarrow &  \sigma_4\\
\end{array}
$$

\subsubsection{Sections of quantum vector bundles over  ${A_{11}}/{E_6}$}

Using the previous table, and using  a formal analogy,
we associate a quantum vector bundle
\footnote{There are several ways to define quantum principal bundles and 
associated quantum vector bundles, 
(see for instance \cite{CGT:bundles}), but we do not use these
technical definitions here.}
 to each point of the $E_6$ graph 
and decompose its spaces of sections $\Gamma_{\sigma_p}$,  using induction,  
exactly as we did in the classical case
(for instance we see that $\sigma_0$ can be obtained {\sl from} the
reduction of $\tau_0$ and $\tau_6$). We obtain:

\begin{eqnarray*}
  \Gamma_{\sigma_0}  & = & \tau_0 + \tau_6 \\
  \Gamma_{\sigma_1}  & = & \tau_1 + \tau_5 + \tau_7  \\
  \Gamma_{\sigma_2} & = & \tau_2+ \tau_4 + \tau_6 + \tau_8 \\
  \Gamma_{\sigma_5}  & = & \tau_3  + \tau_5 + \tau_9 \\
  \Gamma_{\sigma_4}  & = & \tau_4 + \tau_{10} \\
  \Gamma_{\sigma_3} & = & \tau_3 + \tau_7
\end{eqnarray*}

This information can also be displayed as

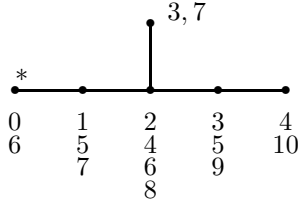
\begin{figure}[h]
\unitlength 0.6mm
\begin{center}
\begin{picture}(95,35)
\thinlines 
\multiput(25,10)(15,0){5}{\circle*{2}}
\put(55,25){\circle*{2}}
\put(25,12){$\ast$}
\thicklines
\put(25,10){\line(1,0){60}}
\put(55,10){\line(0,1){15}}

\put(25,3){\makebox(0,0){$0$}}
\put(25,-2){\makebox(0,0){$6$}}

\put(40,3){\makebox(0,0){$1$}}
\put(40,-2){\makebox(0,0){$5$}}
\put(40,-7){\makebox(0,0){$7$}}

\put(55,3){\makebox(0,0){$2$}}
\put(55,-2){\makebox(0,0){$4$}}
\put(55,-7){\makebox(0,0){$6$}}
\put(55,-12){\makebox(0,0){$8$}}

\put(70,3){\makebox(0,0){$3$}}
\put(70,-2){\makebox(0,0){$5$}}
\put(70,-7){\makebox(0,0){$9$}}

\put(85,3){\makebox(0,0){$4$}}
\put(85,-2){\makebox(0,0){$10$}}

\put(63,27){\makebox(0,0){$3,7$}}
\end{picture}
\bigskip
\label{fig:EssPath0}
\end{center}
\end{figure}

The quantum dimension of 
$\Gamma_{\sigma_0}$ is $$[1]+[7]=[1]+[5]=1 + \frac{\sin(5 \pi/12)}{ 
\sin(\pi/12)}= \frac{2\sqrt 3}{\sqrt 3 -1}$$ 
 Morally this is the  quantum dimension of the space of ``functions'' on the quantum space 
$A_{11}/E_6$.
We may then check that, by dividing the $q$-dimension of each space of sections
$\Gamma_{\sigma_p}$ by the above $q$-dimension of $\Gamma_{\sigma_0}$, we recover
exactly the $q$-dimensions of the ``typical fibres'', \ie the $q$-dimensions
of the irreducible representations $\sigma_p$, already obtained from the normalized
Perron-Frobenius vector associated with the graph $E_6$. We have therefore
a perfect analogy with the classical situation.

\section{Appendix: The general notion of essential paths on a graph $G$}

The following definitions are not needed if we only want to count the 
number of essential paths on a graph. They are necessary if we want to
obtain explicit expressions for them.
The definitions are adapted from \cite{Ocneanu:paths}.

Call $\beta$ the norm of the graph $G$ (the biggest eigenvalue of its 
adjacency matrix ${\cal G}$) 
and  $D_{i}$ the components of the (normalized) Perron Frobenius eigenvector.  
Call $\sigma_{i}$ the vertices of $G$ and, if 
$\sigma_{j}$ is a neighbour of 
$\sigma_{i}$, call $\xi_{ij}$ the oriented edge
from $\sigma_{i}$ to $\sigma_{j}$. If $G$ is unoriented (the case for $ADE$
and affine $ADE$ diagrams), each edge should be considered  as carrying
both orientations.

An elementary path can be written either as a finite 
sequence of consecutive (\ie neighbours on the graph) vertices, 
$[\sigma_{a_1} \sigma_{a_2} \sigma_{a_3} \ldots ]$,
or, better, as a sequence $(\xi(1)\xi(2)\ldots)$ of consecutive edges, with
$\xi(1) = \xi_{a_{1}a_{2}}= \sigma_{a_1} \sigma_{a_2} $,
$\xi(2) = \xi_{a_{2}a_{3}} = \sigma_{a_2}  \sigma_{a_3} $, \etc.
Vertices are considered as paths of length $0$.

The length of the (possibly backtracking) path $( \xi(1)\xi(2)\ldots 
\xi(p) )$ is $p$.
We call $r(\xi_{ij})=\sigma_{j}$, the range of $\xi_{ij}$
and $s(\xi_{ij})=\sigma_{i}$, the source of $\xi_{ij}$.

For all edges $\xi(n+1) = \xi_{ij}$ that appear in an elementary path, 
we set  ${\xi(n+1)}^{-1} \doteq \xi{ji}$.

For every integer $n >0$, the annihilation operator $C_{n}$,
acting on elementary paths of length $p$ is defined
as follows:  if $p \leq n$, $C_{n}$ vanishes, whereas if $ p \geq  n+1$ then
$$
C_{n} (\xi(1)\xi(2)\ldots\xi(n)\xi(n+1)\ldots) = 
\sqrt\frac{D_{r(\xi(n))}}{D_{s(\xi(n))}} 
\delta_{\xi(n),{\xi(n+1)}^{-1}}
 (\xi(1)\xi(2)\ldots{\hat\xi(n)}{\hat\xi(n+1)}\ldots) 
$$
Here, the symbol ``hat'' ( like  in $\hat \xi$) denotes omission.
The result is therefore either $0$ or a linear combination of paths of length $p-2$.
Intuitively, $C_{n}$ chops the round trip that possibly appears
at positions $n$ and $n+1$.

A path is called essential if it belongs to 
the intersection of the kernels 
of the anihilators $C_{n}$'s.

In the case of the diagram $E_{6}^{(1)}$, for instance
$$C_{3}(\xi_{12}\xi_{23}\xi_{32'}\xi_{2'3}\xi_{32}) = 
C_{3}(\xi_{12}\xi_{23}\xi_{32''}\xi_{2''3}\xi_{32})
 = \sqrt{\frac{2}{3}}(\xi_{12}\xi_{23}\xi_{32})$$

The following difference of non essential paths of length $5$ starting at $\sigma_{1}$ and ending at $\sigma_{2}$
is an essential path of length $5$  on $E_{6}^{(1)}$:
$$(\xi_{12}\xi_{23}\xi_{32'}\xi_{2'3}\xi_{32}) - (\xi_{12}\xi_{23}\xi_{32''}\xi_{2''3}\xi_{32})
= [1,2,3,2',3,2]-[1,2,3,2'',3,2]$$

Another example: in the case of the diagram $E_{6}$ (square brackets
enclose $q$-numbers),  
\begin{eqnarray*}
C_{3}(\xi_{01}\xi_{12}\xi_{23}\xi_{32}) &=&  \sqrt \frac {[1]}{[2]} 
(\xi_{01}\xi_{12} \\
C_{3}(\xi_{01}\xi_{12}\xi_{25}\xi_{52}) &=&  \sqrt \frac {[2]}{[3]} 
(\xi_{01}\xi_{12}
 \end{eqnarray*}

The following difference of non essential paths of length $4$ starting at $\sigma_{0}$ and ending at $\sigma_{2}$
is an essential path of length $4$  on $E_{6}$:
$$\sqrt{[2]} (\xi_{01}\xi_{12}\xi_{23}\xi_{32})  - \sqrt \frac{[3]}{[2]} (\xi_{01}\xi_{12}\xi_{25}\xi_{52}) 
= \sqrt{[2]}  [0,1,2,3,2]- \sqrt \frac{[3]}{[2]}[0,1,2,5,2]$$
Remember the values of the $q$-numbers: $[2] = \frac{\sqrt 2}{\sqrt 3 -1}$ and $[3] = \frac{2}{\sqrt 3 -1}$. 

Acting on elementary path of length $p$, the creating operators $C^{\dag}_{n}$ are defined as follows:
if $n > p+1$, $C^{\dag}_{n}$ vanishes and, if $n \leq p+1$ then,
setting $j = r(\xi(n-1))$, 
$$
C^{\dag}_{n} (\xi(1)\ldots\xi(n-1)\ldots) = \sum_{d(j,k)=1}
\sqrt(\frac{D_{k}}{D_{j}})  (\xi(1)\ldots\xi(n-1)\xi_{jk}\xi_{kj}\ldots)
$$
The above sum is taken over the neighbours $\sigma_{k}$ of $\sigma_{j}$ on the graph.
Intuitively, this operator adds one  (or several) small round trip(s) 
at position $n$. 
The result is therefore either $0$ or a linear combination of paths of 
length $p+2$.

For instance, on paths of length zero (\ie vertices),
$$
C^{\dag}_{1} (\sigma_{j}) = \sum_{d(j,k)=1}
\sqrt(\frac{D_{k}}{D_{j}}) \xi_{jk}\xi_{kj} = \sum_{d(j,k)=1}
\sqrt(\frac{D_{k}}{D_{j}}) \, [\sigma_{j}\sigma_{k}\sigma_{j}]
$$

Jones' projectors $e_{k}$ are defined (as endomorphisms of 
$Path^p$) by 
$$
e_{k} \doteq \frac{1}{\beta} C^{\dag}_{k} C_{k} 
$$

The reader can check that all Jones-Temperley-Lieb relations 
between the $e_i$ are satisfied.
Essential paths can also be defined as elements of the intersection of the 
kernels of the Jones projectors $e_{i}$'s.

\end{document}